\documentstyle[psfig]{l-aa}
\newcommand{\etal}{{\it et al. }}
\newcommand{\kms}{km~s$^{-1}~$}

\newcommand{\mpm}{$\pm$}

\begin{document}
\thesaurus{06           
          (12.02.1;     
           12.03.1;     
           12.26.1;     
           19.12.1;     
           19.16.1;     
           19.48.1)     
          }
\title{Spectral line identification of the ultraviolet spectrum
of the peculiar post-AGB star HD~101584 (F0Iape)}
\title{The ultraviolet spectrum
of the peculiar post-AGB star HD~101584}
\author{Eric J. Bakker}
\institute{SRON Laboratory for Space Research \\
           Sorbonnelaan 2 \\
           NL-3584 CA Utrecht \\
           The Netherlands }
\date{received March 12 1993, accepted june 13 1993}
\maketitle

\begin{abstract}
The high-resolution ($R \approx 10^{4}$) near-ultraviolet IUE spectrum
of the peculiar post Asymptotic Giant Branch (post-AGB) supergiant
HD~101584 (F0Iape) has been analyzed. In the
wavelength range between 2500 and 3000~\AA, 291 out of 332
absorption features are identified. The main contribution comes
from FeII, CrII and MnII.
Most of the absorption features are blends from several
absorption lines. This is mainly due to the intrinsic broad line
profiles in the spectrum of HD~101584 in combination with the
many absorption lines which occur in the UV for a F-type spectrum.

The list compiled gives for all measured absorption features
between   2500 and  3000~\AA~ the measured
central wavelength of the core of the profile, the equivalent width
of the profile if this could be determined, the depth of the profile,
an asymmetry factor, a quality factor and lists  the
transitions which are responsible for the feature.
An analysis of the radial velocities of the absorption profiles
will be give in a separate paper.
\end{abstract}

\section{Introduction}

The peculiar supergiant HD~101584 has been classified as a F0Iape
star (Hoffleit~\etal \cite{art1hoffleit}). The star is associated
with a strong infrared source
(Humpreys \& Ney \cite{art1humphreys:ney:a} \cite{art1humphreys:ney:b};
Parthasarathy \&
Pottasch \cite{art1parthasarathy:pottasch}), which is normally interpreted as
a combination of cold and hot dust. HD~101584 does also show several
millimeter CO($J=1 \rightarrow 0$)  emission features
(Trams~\etal \cite{art1trams:veen};
Loup~\etal \cite{art1loup:froveille}) and outwards accelerated
bipolar OH maser emission (Te Lintel Hekkert~\etal \cite{art1lintel:chapman}).
Radial velocity measurements of absorption lines show variations
which are not well understood yet. It is not evident that these
variations are due to binarity of the system.

\begin{table}
\caption{Data on HD~101584 and $\alpha$~Lep}
\label{art1tab-star}
\centerline{\begin{tabular}{lll}
\hline
                 &  HD~101584                & HD~36673                 \\
                 &                           & $\alpha$~Lep             \\
\hline
                 &                           &                          \\
$\alpha  [2000]$ &$ 11^{h}\,40^{m}\,58.8^{s}$& $05^{h}\,32^{m}\,43.7^{s}$\\
$\delta  [2000]$ &$-55^{o}\,34^{'}\,25^{''}$ & $-17^{o}\,49^{'}\,20^{''}$\\
$b^{II}$         &$+5.94^{o}$                & $-25.14^{o}$             \\
$l^{II}$         &$293.03^{o}$               & $220.95^{o}$             \\
Spectral Type    &F0Iape                     &  F0Ib                    \\
App. Visual Magn.&7.01                       &  2.58                    \\
$B-V$            &0.39                       &  0.21                    \\
$v \sin i$ [\kms]&                           & 15                       \\
                 &                           &                          \\
\hline
\end{tabular}}
\centerline{Information based on Hoffleit (\cite{art1hoffleit})}
\end{table}

Humpreys (\cite{art1humphreys}) suggested that
the object is a binary system with a period of 3.5 years.
In this model the strong IR excess is attributed to a
secondary
cool M-type star which fills its Roche lobe. The primary,
the observed star, is a hot
accreting A (or B) star which accretes gas and dust form the secondaries
Roche lobe. The A-type star is thus observed with a F-type like spectrum.
The mass-loss from the secondary produces a common envelope around the system
which give a P-Cygni profile of e.g. the H$\alpha$ line.

An alternative model for this object is by Parthasarathy and Pottasch
(\cite{art1parthasarathy:pottasch}).
They suggest that HD~101584 is a transition object form the AGB
to the planetary nebula stage, the
post Asymptotic Giant Branch (post-AGB) phase.
The large infrared excess can be explained
by the combination of a hot and a cold dust shell
of  $750$ and $120$~K respectively. A study by Trams~\etal
(\cite{art1trams:waters}) classifies HD~101584 as a post-AGB star
on basis of its infrared excess and galactic latitude.

An extension of the post-AGB theory has been made by Trams~\etal
(\cite{art1trams:veen}).
They discuss the millimeter CO($J=1 \rightarrow 0$) emission features
of HD~101584 in terms of multiple ejections
of shells, or the diametrical ejections of blobs. The CO peak at the stellar
rest velocity, and the narrow stationary optical emission lines could
originate from a flattened envelope (or disk).

The aim of this study is to investigate this interesting star in order
to gain insight in the evolution of post-AGB binaries and
to extend our current knowledge of HD~101584 from the radio, millimeter,
infrared and optical to the ultraviolet by making a UV spectral line
identification. The compiled list (Table~\ref{art1tab-list})
is an absorption line identification between
2500~\AA~ and 3000~\AA~ based on IUE spectra.
This list will be subject to further study on the radial velocities
of the identified lines and on the asymmetry of the line
profiles measured.

\begin{figure*}
\centerline{\hbox{\psfig{figure=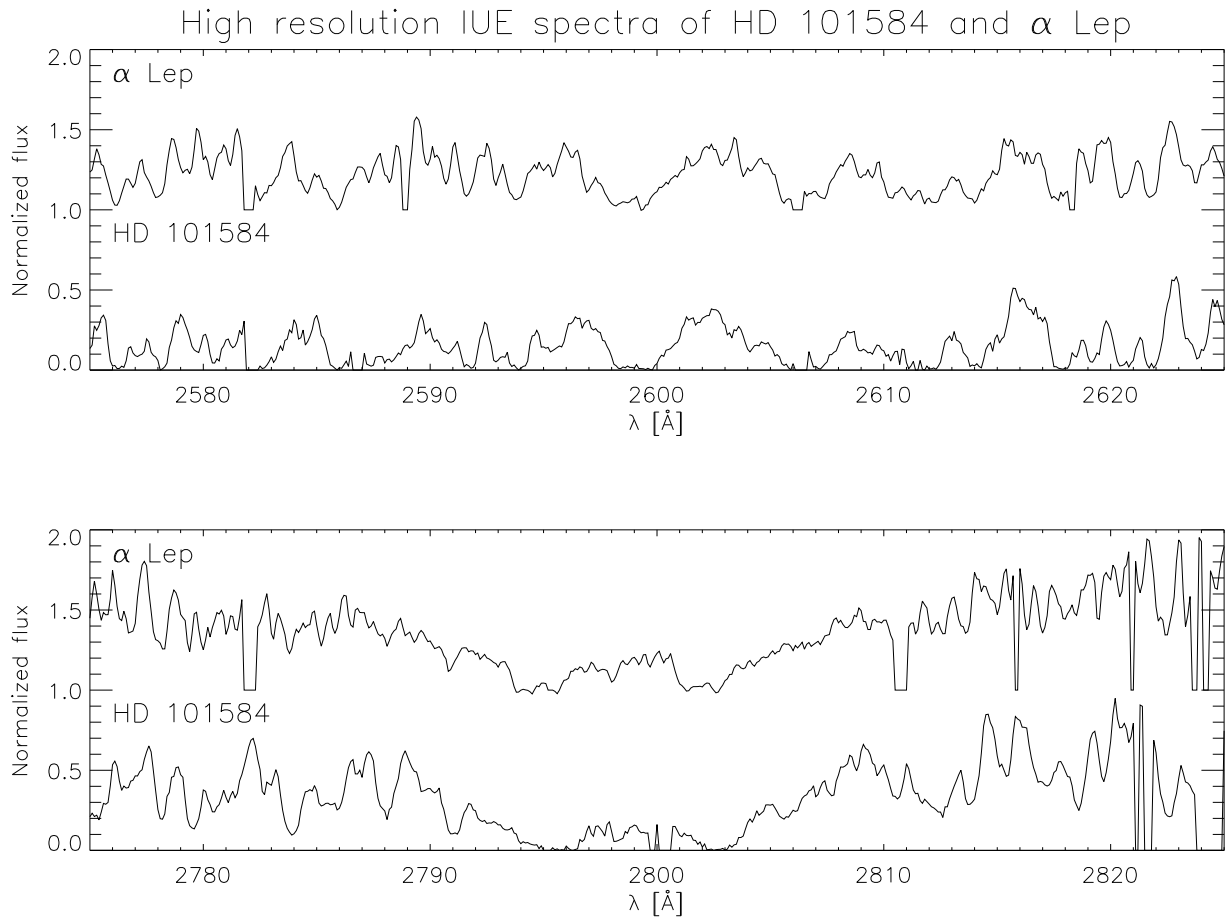,width=\textwidth}}}
\caption{The ultraviolet spectrum of the peculiar supergiant HD~101584
           (LWP17369)
           and of the normal supergiant $\alpha$~Lep (LWR5781).
           The absorption lines
           in the spectrum of HD~101584 are intrinsically broader than the
           corresponding absorption lines in $\alpha$~Lep.
           The normalized flux of $\alpha$~Lep is shifted by $+1$
           to allow a better comparison with HD~101584. The spectra are
           not heliocentrically corrected in this plot}
\label{art1fig-hires}
\end{figure*}

The spectra used are obtained with the International
Ultraviolet Explorer (IUE).
The 1900 to 3200~\AA~ UV band
has been detected in  high-resolution mode ($ R \approx 10^{4}$ ),
which gives a velocity resolution of approximately 6~\kms
(Kondo \cite{art1kondo}).
For wavelengths short wards of 2500~\AA~ the moderate
interstellar and circumstellar extinction, in combination with the
decrease of the continuum level,
give a too low signal-to-noise ratio to make a reliable identification
possible. For wavelengths long wards of 3000~\AA~ the echelle orders
do not fully overlap
and this introduces blind spots in the spectrum. For these two reasons
this study limits itself
to the line identification in the spectral range between
2500 and 3000~\AA.
By comparing
the spectrum of HD~101584 with a
standard F-type supergiant  $\alpha$~Lep (F0Ib),
a line identification is made on basis of a multiplet fit technique
as described in Sect.~\ref{art1sec:ident}. By a first comparison of
the spectrum of HD~101584 with the comparison star $\alpha$~Lep it is
evident that the absorption features in HD~101584 are intrinsically
broader than the corresponding features in $ \alpha$~Lep
(Fig.~\ref{art1fig-hires}).
This effect accounts for the numerous strongly blended features and
a poorly defined continuum level of the spectrum.

The result of this study is a list (Table~\ref{art1tab-list})
of 332 absorption features
of the object HD~101584 in the wavelength region between
2500  and  3000~\AA.
We also provide
information on the observed central wavelength, equivalent width,
depth of the feature with respect to the adopted local continuum,
an asymmetry factor,
and the possible contribution of various blends to the absorption features.
The data for the absorption lines are from
Moore's (\cite{art1moore50} and \cite{art1moore62})
ultraviolet multiplet tables.

\section{The observations}

The observations discussed in this paper are high-resolution
long wavelength spectra, LWP17369, LWR4052 and LWR5781
made with the International Ultraviolet
Explorer (IUE) as described in Kondo (\cite{art1kondo}).
The observational parameters of the spectra are given in
Table~\ref{art1tab-star} and \ref{art1tab-info} and
the raw IUE image were processed with the Starlink program IUEDR
(version 1.3). The extracted spectra
were used for this study. Due to the lack of clearly
detectable interstellar
absorption lines is was not possible to make an absolute wavelength
calibration based on interstellar absorption lines. For LWP17369
we used the
heliocentric velocity correction from the IUE software. For
LWR5042 and LWR5781 no heliocentric velocity corrected were available
from the IUE log file and
we corrected for the earth motion and neglected
the additional satellite velocity. This introduces a systematic velocity
shift of one spectrum to another with a maximum velocity of
3.3~\kms.
There was no attempt made to reprocess the reference
spectrum of $\alpha$~~Lep.

\begin{table*}
\caption{Information on the spectra of HD~101584}
\label{art1tab-info}
\centerline{\begin{tabular}{llll}
\hline
                       & HD~101584      & HD~101584    & $\alpha$~Lep \\
\hline
                       &                &              &                    \\
Image                  & LWP17369       & LWR5042      & LWR5781            \\
Observation date       & 15 Febr. 1990  & 14 July 1979 & 10 Aug. 1979       \\
Aperture               & large          & large        & small              \\
Integration time       & 140 minutes    & 120 minutes  & 10 minutes         \\
Ground Station         & Vilspa (Spain )& Goddard      & Goddard            \\
Reprocessed            & IUEDR 1.3      & IUEDR 1.3    & 10 Aug. 1979       \\
Heliocentric corrected & Yes            & Yes          & Yes                \\
                       &                &              &                    \\
\hline
\end{tabular}}
\end{table*}

\section{The line identification}
\subsection{The method}
\label{art1sec:ident}

The high-resolution near-ultraviolet spectrum of the supergiant HD~101584
is characterized by intrinsically broad absorption
lines (Fig.~\ref{art1fig-hires}).
This results in a spectrum containing numerous strongly
blended absorption features and a not well determined continuum
level. By comparison of the spectrum of HD~101584 with a standard
F-type supergiant, $\alpha$~Lep (F0Ib),
the line identification is simplified.
Since the IUE spectrum of HD~101584 has
intrinsically broad absorption lines,
it seems almost useless to try to make an
identification by means of some kind of automatic computer
algorithm. The spectral identification described in this work
is therefore based on individual multiplet fitting.
The identification is made for the IUE spectrum LWP 17369 with LWR 5042
to fill in the missing
data points of the former spectrum. The missing data points are due
to reseau marks and wavelength gaps between two successive echelles.
The absorption features measured from this second spectrum are marked with
an asterisk ($\ast$) in Table~\ref{art1tab-list}.
The data of these features should not be mixed
with the data of features from the other spectrum, because there seem to be
absorption line dependent velocity and intensity variations.

{}From the literature (Van der Hucht~\etal \cite{art1hucht:lamers}) a first
estimate about the multiplets
which makes the largest contributions to the broad features
could be obtained. Starting with the most
conspicuous multiplets, FeII 60, 61, 62 and 63, all lines of a
given multiplet were checked to see if there was a corresponding absorption
feature in the spectrum near the laboratory wavelength for this line.
A multiplet was designated ``detected'' if most of the multiplet lines
were measured with an equivalent width ratio in agreement with the
intensity values as given in the multiplet table and a consistent
Doppler velocity of the lines within one multiplet.
However due to the large number of blended lines in the spectrum, it was
not always evident that a multiplet was detected and the decision
was based on common sense.

The multiplet fitting technique was finished when most of the
absorption features were identified.
The remaining features were identified
by looking up the apparent central wavelength of the absorption feature
in the finding list. Possible candidate lines were selected on basis of
a multiplet fit. Some features remained unidentified. By looking up these
features in the Moore~\etal (\cite{art1moore82})
solar spectrum atlas, some of them could be identified.

This method of line identification makes use of only a limited number of
multiplets. Using the foreknowledge of previous line identifications
on F-type supergiant spectra, one
limits oneself by checking only the multiplets which have a high
change of being detected. This means that low abundance elements are
not checked, nor higher ionization degrees than expected for a
F-type spectrum, nor those multiplet which have a high excitation level.
The disadvantage of this technique, that only the expected ions
will be tested, and
that anomalies are not recognize, is lifted by  introducing
a comparison star. The spectrum of HD~101584
only shows small differences with the reference spectrum of
$\alpha$~Lep.
The most prominent difference is the intrinsic broad lines profiles
in the spectrum, but there seems to be no significant different
abundance pattern, nor additional lines.

The final absorption feature line-identification list of the ultraviolet
spectrum (2500-3000~\AA) of HD~101584
is given in Table~\ref{art1tab-list}.
It lists
the main characteristics of the features, possible candidate absorption lines,
and gives the actual identification made in this study. The
heliocentric velocities derived from the apparent central wavelength of the
feature and
the laboratory wavelength are also given.

\subsection{Description of the tables}

The format of the table is as follow:
\begin{enumerate}
\item  $\lambda_{\rm obs.}$ [\AA]:
       the observed central wavelength of the absorption feature
\item  $W$ [\AA]: the observed equivalent width.
       The error on the given equivalent
       width is about 25\% which is mainly due to the lack of a well
       defined continuum level.
       If one of the wings of the profile
       could not be measured, the equivalent width is calculated assuming
       a symmetric profile.
       If both wings could not be measured
       this space is left blank
\item  $D$ [\%]: the depth of the profile with respect to the
       adopted local continuum. This
       value ranges from 100, very strong absorption, to almost 0, for
       very weak absorption
\item  $A$: the asymmetry factor defined as the ratio of the red over blue
       equivalent width. A symmetric profile has an asymmetry factor of
       one. For features which have an excess of red absorption, the
       asymmetry factor is greater than one. For features with extra
       blue absorption the asymmetry factor is smaller than one.
       If one of the wings could not be measured a code is given:
       a ``R B'' (red blended) or ``B B'' (blue blended)
       for the lack of a measurement of the red or blue wing respectively
\item  $Q$: the quality factor for the identification and the
       absorption
       feature. If the quality factor is ``3'' the line is positively
       identified with one absorption line. Blends of this line
       are of minor importance. The lines marked with a ``3'' can
       be used for further study on the velocities and asymmetry of
       the line. A quality of ``2''
       means that either the line is weakly blended,
       or the same absorption feature in $\alpha$~Lep is split into two
       components, or there is more than one good candidate absorption
       line. A quality of ``1'' is an unreliable line for further study.
       The identification made is however correct, but there are strong
       blends, or the line is very weak. If the quality factor is
       followed by an asterisk the data of the absorption feature is
       from the additional spectrum (LWR5042)
\item  $ \lambda_{\rm lab.}$ [\AA]:
       the laboratory wavelength of the candidate line
\item  ion(mult): the element and multiplet number for the given
       absorption line
\item  $\chi$ [eV]: lower level excitation energy of transition
       which give the candidate absorption line
\item  $C$: contribution of the blends to the observed line.
       A candidate absorption line marked with
       a ``{\bf P}'' is the {\bf primary contributor} to the feature.
       ``S'' the secondary, and ``M''
       for lines which give only a minor contribution to the features.
\item  $v$ [km~s$^{-1}$]: radial heliocentric velocity for identified
       absorption feature
\item  remarks: remarks on the line identification.
       ``U'' stands for unidentified feature
\end{enumerate}

\section{Discussion}

In the wavelength range between 2500  and  3000~\AA, 332
absorption features have been measured and tabulated in
Table~\ref{art1tab-list}.
291 of the features could be identified. Most of them have only one
primary contributor. An overview of all transitions
which are marked as primary contributor is given in
Table~\ref{art1tab-contrib}.
As expected the spectrum is strongly dominated by FeII, which is the
main degree of ionization of Fe in the spectrum of HD~101584. It seems
that there are also some lines from SiI and TiII in the spectrum.

\begin{table}
\caption{An overview of the ions marked as principal contributor to
           an absorption feature in the UV spectrum of HD~101584.
           Of the 332 features, six have two primary contributors,
           the others only one. The number of occurrences as
           primary contributor in the identification list}
\label{art1tab-contrib}
\centerline{\begin{tabular}{ll|ll}
\hline
Ion             & No. & Ion  & No.  \\
\hline
                &    &      &     \\
Unidentified    & 41 &      &     \\
Mg~I            & 8  & Cr~I & 1   \\
MgII            & 7  & CrII & 62  \\
Si~I            & 4  & MnII & 47  \\
TiII            & 4  & Fe~I & 12  \\
V~II            & 19 & FeII & 133 \\
                &    &      &     \\
\hline
\end{tabular}}
\end{table}

The 41 unidentified features are tabulated
in Table~\ref{art1tab-unident}. For seven of these features a possible
identification could be made as given in Table~\ref{art1tab-ident}.

\begin{table}
\caption{Unidentified features in the spectrum of HD~101584}
\label{art1tab-unident}
\centerline{\begin{tabular}{lllll}
\hline
\multicolumn{5}{c}{$\lambda_{\rm obs}$ [\AA]} \\
\hline
        &         &         &         &         \\
2508.14 & 2616.00 & 2705.05 & 2807.06 & 2901.52 \\
2508.69 & 2637.00 & 2716.25 & 2808.81 & 2904.42 \\
2522.09 & 2641.50 & 2735.16 & 2818.56 & 2966.37 \\
2552.44 & 2642.35 & 2737.26 & 2819.76 & 2968.57 \\
2553.89 & 2649.75 & 2770.86 & 2820.81 & 2983.67 \\
2557.49 & 2655.25 & 2773.06 & 2833.76 &         \\
2584.49 & 2683.10 &         & 2834.61 &         \\
2587.34 & 2686.25 &         & 2836.56 &         \\
2589.09 &         &         & 2839.26 &         \\
2592.84 &         &         & 2845.71 &         \\
        &         &         & 2884.22 &         \\
        &         &         & 2888.30 &         \\
        &         &         &         &         \\
\hline
\end{tabular}}
\end{table}

\begin{table}
\caption{Possible identification of unidentified features}
\label{art1tab-ident}
\centerline{\begin{tabular}{ll}
\hline
$\lambda_{\rm obs}$ [\AA]& Poss.      \\
                         & ident.     \\
\hline
                         &            \\
2589.09                  & MnII       \\
2616.00                  & Fe~I       \\
2683.10                  & V~II (I.S.)\\
2735.16                  & ZrII       \\
2820.81                  & OH         \\
2834.61                  & Fe~I       \\
2983.67                  & NiII       \\
                         &            \\
\hline
\end{tabular}}
\end{table}

Table~\ref{art1tab-list} gives use the opportunity to look at the statistics
on the velocities and on the asymmetry  factor.

\begin{figure*}
\centerline{\hbox{\psfig{figure=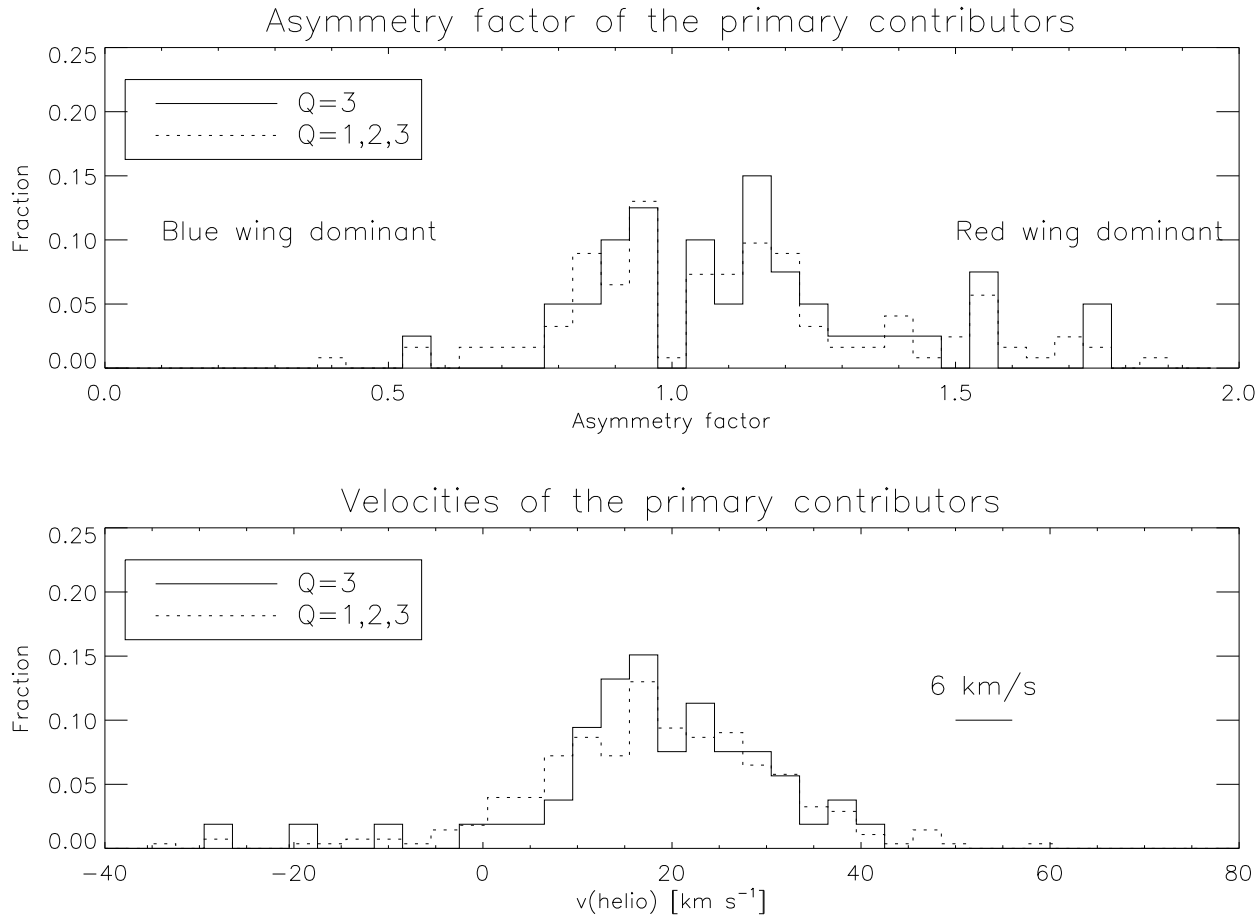,width=\textwidth}}}
\caption{The upper histogram shows a numbercount on the asymmetry
           factor. It is noted that the factor does not spread around 1
           but rather on a value a little bit higher.
           The lower histogram is a numbercount on the velocities.
           The IUE resolution predicts an accuracy of 6~\kms,
           significant smaller then measured in this spectrum.
           The histograms are made for all transitions marked with a ``P''
           and all quality factors (dashed line).
           The solid lines are made by using only the quality $Q=3$ lines.
           The histograms are normalized on 1 to allow comparison}
\label{art1fig-statvel1}
\end{figure*}

\begin{figure*}
\centerline{\hbox{\psfig{figure=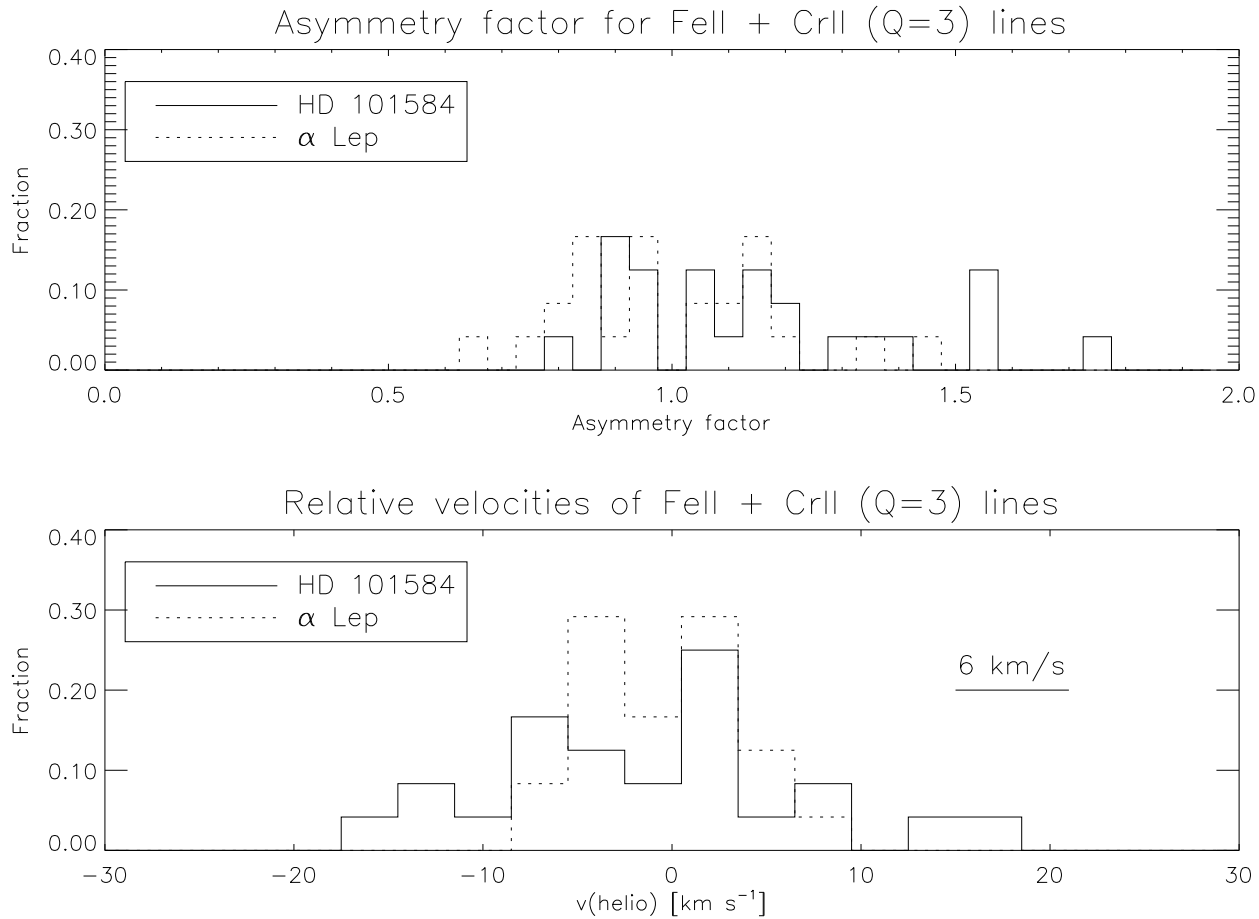,width=\textwidth}}}
\caption{The upper histogram shows a numbercount on the asymmetry
           factor, the lower on the velocities relative to the average
           velocity. In this figure only the FeII and CrII lines with a quality
           factor of $Q=3$
           are used and which occur in the spectra of both stars.
           It is noted that the asymmetry factor for HD~101584 is on the
           average larger than for $\alpha$~Lep, and that the asymmetry
           factor
           for HD~101584 is not centered around 1. The velocity spread in
           $\alpha$~Lep is as expected from the IUE resolution. The
           spread on the same lines in HD~101584 is significant larger than
           expected.}
\label{art1fig-statvel2}
\end{figure*}

\subsection{Asymmetric line profiles}

Fig.~\ref{art1fig-statvel1} shows the number of profiles within a given
range of asymmetry factors. There is a tendency for
an  asymmetry factor larger than one.
This means that the slope of the red wing of the profile is less steep
than the blue wing. This is unexpected because for
photospheric lines one expect unity, and for wind line a value smaller
than one.

By making a comparison of HD~101584 with $\alpha$~Lep
using only those lines
which have a high quality (FeII and CrII with $Q=3$) it is clear that
the asymmetry factor in HD~101584 is larger than for $\alpha$~Lep
(Fig.~\ref{art1fig-statvel2}).

\subsection{Large velocity spread in measured Doppler velocities}

{}From Kondo  \cite{art1kondo}
a velocity spread from \mpm6~\kms based on the IUE resolution
is expected, it is however clear from Fig.\ref{art1fig-statvel1} that
the real spread in measured Doppler velocities is much larger for
the lines which are marked as primary contributor to an absorption
feature. A study on this interesting problem is in progress.
(Bakker \cite{art1bakker}).

By making a comparison of HD~101584 with $\alpha$~Lep
using only those lines
which have a high quality (FeII and CrII with $Q=3$) it is clear that
the velocity spread in HD~101584 is larger than for $\alpha$~Lep.

\begin{figure}
\centerline{\hbox{\psfig{figure=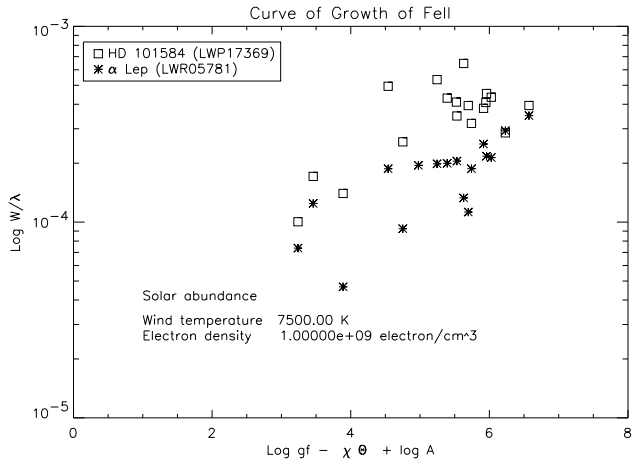,width=\columnwidth}}}
\caption{The Curve of Growth for $\alpha$~Lep  and HD~10158.
           Only the Fe~II lines with a quality factor $Q=3$ are used in this
           graph. The data points for HD~101584 are higher in
           comparison to $\alpha$~Lep. This shows the presence
           of an extreme broadening mechanism of the absorption profiles.
           The data points of $\alpha$~Lep seem to follow a
           standard COG curve, which imply that the $Q=3$ annotation is
           a strong one}
\label{art1fig-cog}
\end{figure}

\subsection{Curve of Growth for FeII}

To see whether the quality factor has any significance, a curve of
growth (COG) for the FeII identification with $Q=3$ in $\alpha$~Lep
has been made (Fig.~\ref{art1fig-cog}). It is evident from the
figure that the $Q=3$ annotation is a strong one, although there
are data points for $\alpha$~Lep which seem to have a too low
equivalent width. The COG of the same identified line of FeII for HD~101584
is superimposed in this figure. The horizontal part of the COG for HD~101584
is significantly shifted up due to a broadening mechanism of the
absorption features, whereas the spread in the data points is also larger
than for the reference star.

\acknowledgements{The author wants to thank Henny Lamers, Rens Waters
and Christoffel Waelkens for the many stimulating
and constructive discussions on this work, Norman Trams for making the
ultraviolet spectrum of HD~101584 available for this study.
Lex Kaper is thanked for the
help with the IUEDR data reduction package.
The author was supported by grant no. 782-371-040 by ASTRON,
which receives funds from the Netherlands Organization for
the Advancement of Pure Research (NWO).
This research has made use of the Simbad database, operated at
CDS, Strasbourg, France.}

\clearpage
\begin{table*}
\caption{UV line identification of HD~101584 (2500 to  3000~\AA)}
\label{art1tab-list}
\begin{small}
\centerline{
\begin{tabular}{|llllllllllp{25mm}|}
\hline
$\lambda_{\rm obs}$&$W$&$D$&$A$ &$Q$&$\lambda_{\rm lab}$
&ion(mult) &$\chi$ &$C$ &$v$ &remark\\
{}~[\AA]&[\AA]     & [\%]     &  &  &~[\AA] &   &[eV]   &
&\multicolumn{2}{l|}{[\kms]}   \\
\hline
2501.14& 0.54&  94&     & 1  &2500.919 &FeII(357)& 4.71&   P&    27&
                                   \\
\cline{6-6}
2502.69&     &  94&     & 1  &2502.388 &FeII(207)& 3.21&   P&    36&
                                   \\
\cline{6-6}
2505.34& 0.87& 100& 1.60& 2  &2505.217 &FeII(033)& 0.30&   P&    15&
                                   \\
\cline{6-6}
2508.14&     &  76&     & 2  &         &         &     &   P&      & U
                                   \\
\cline{6-6}
2508.69&     &  73&     & 2  &         &         &     &   P&      & U
                                   \\
\cline{6-6}
2509.34& 1.24&  87& 0.91& 3  &2509.117 &FeII(242)& 3.23&   P&    27&
                                   \\
\hline
\end{tabular}}
\end{small}
\end{table*}

\clearpage
\begin{table*}
\addtocounter{table}{-1}
\caption{continued}
\begin{small}
\centerline{\begin{tabular}{|llllllllllp{24mm}|}
\hline
$\lambda_{\rm obs}$&$W$        &$D$      &$A$
&$Q$                &
$\lambda_{\rm lab}$&ion(mult) &$\chi$ &$C$
&$v$                &remark\\
{}~[\AA]&[\AA]     & [\%]
 &  &                 &
{}~[\AA] &            &[eV]   &
                     &\multicolumn{2}{l|}{[\kms]}   \\
\hline
       &     &    &     &    &2556.893 &MnII(020)& 3.41&   S&      &
                                   \\
2557.19&     &  92&     & 1  &2557.079 &FeII(158)& 2.68&   P&    13&
                                   \\
\cline{6-6}
2557.49&     &  90&     & 1  &         &         &     &   P&      & U
                                   \\
       &     &    &     &    &2557.45  &CrII(089)& 3.74&   M&      &
                                   \\
\cline{6-6}
2558.69& 0.56&  81&     & 2  &2558.604 &MnII(020)& 3.40&   P&    10&
                                   \\
\cline{6-6}
2559.64&     &  95&     & 1  &2559.415 &MnII(020)& 3.41&   P&    26&
                                   \\
\hline
\end{tabular}}
\end{small}
\end{table*}

\clearpage
\begin{table*}
\addtocounter{table}{-1}
\caption{continued}
\begin{small}
\centerline{\begin{tabular}{|llllllllllp{24mm}|}
\hline
$\lambda_{\rm obs}$&$W$        &$D$      &$A$
&$Q$                &
$\lambda_{\rm lab}$&ion(mult) &$\chi$ &$C$
&$v$                &remark\\
{}~[\AA]&[\AA]     & [\%]
 &  &                 &
{}~[\AA] &            &[eV]   &
                     &\multicolumn{2}{l|}{[\kms]}   \\
\hline
2604.15&     &  79&     & 2  &2603.721 &MnII(036)& 3.69&   P&    49& too strong
                                   \\
\cline{6-6}
       &     &    &     &    &2605.416 &FeII(204)& 3.22&   S&      &
                                   \\
2605.75&     & 100&     & 1  &2605.697 &MnII(001)& 0.00&   P&     6&
                                   \\
\cline{6-6}
       &     &    &     &    &2607.06  &CrII(087)& 3.75&   M&      &
                                   \\
2607.15&     & 100&     & 1  &2607.086 &FeII(001)& 0.08&   P&     7&
                                   \\
\cline{6-6}
       &     &    &     &    &2608.80  &CrII(087)& 3.72&   M&      &
                                   \\
\hline
\end{tabular}}
\end{small}
\end{table*}

\clearpage
\begin{table*}
\addtocounter{table}{-1}
\caption{continued}
\begin{small}
\centerline{\begin{tabular}{|llllllllllp{24mm}|}
\hline
$\lambda_{\rm obs}$&$W$        &$D$      &$A$
&$Q$                &
$\lambda_{\rm lab}$&ion(mult) &$\chi$ &$C$
&$v$                &remark\\
{}~[\AA]&[\AA]     & [\%]
 &  &                 &
{}~[\AA] &            &[eV]   &
                     &\multicolumn{2}{l|}{[\kms]}   \\
\hline
2639.90& 0.70&  77& 1.00& 1  &2639.560 &FeII(221)& 3.32&   P&    39& double
peaked                                 \\
       &     &    &     &    &2639.850 &MnII(052)& 4.05&   M&      &
                                   \\
\cline{6-6}
       &     &    &     &    &2641.124 &FeII(144)& 2.69&   M&      &
                                   \\
2641.50& 0.35&  65& 1.22& 3  &         &         &     &   P&      & U
                                   \\
\cline{6-6}
      &     &    &     &    &2642.015 &FeII(309)& 3.93&   M&      &
                                   \\
2642.35& 0.56&  76& 0.63& 3  &         &         &     &   P&      & U
                                   \\
\hline
\end{tabular}}
\end{small}
\end{table*}

\clearpage
\begin{table*}
\addtocounter{table}{-1}
\caption{continued}
\begin{small}
\centerline{\begin{tabular}{|llllllllllp{24mm}|}
\hline
$\lambda_{\rm obs}$&$W$        &$D$      &$A$
&$Q$                &
$\lambda_{\rm lab}$&ion(mult) &$\chi$ &$C$
&$v$                &remark\\
{}~[\AA]&[\AA]     & [\%]
 &  &                 &
{}~[\AA] &            &[eV]   &
                     &\multicolumn{2}{l|}{[\kms]}   \\
\hline
2683.10& 0.95&  89& 1.18& 2  &         &         &     &   P&      & U; V II IS
line ?                             \\
\cline{6-6}
2685.10& 1.02&  84& B B & 1  &2684.940 &FeII(201)& 3.14&   P&    18& too strong
                                   \\
       &     &    &     &    &2685.19  &CrII(085)& 3.72&   M&      &
                                   \\
       &     &    &     &    &2685.19  &CrII(085)& 3.73&   M&      &
                                   \\
\cline{6-6}
2686.25&     &  76&     & 1  &         &         &     &   P&      & U
                                   \\
       &     &    &     &    &2686.100 &FeII(202)& 3.22&   M&      &
                                   \\
\hline
\end{tabular}}
\end{small}
\end{table*}

\clearpage
\begin{table*}
\addtocounter{table}{-1}
\caption{continued}
\begin{small}
\centerline{\begin{tabular}{|llllllllllp{24mm}|}
\hline
$\lambda_{\rm obs}$&$W$        &$D$      &$A$
&$Q$                &
$\lambda_{\rm lab}$&ion(mult) &$\chi$ &$C$
&$v$                &remark\\
{}~[\AA]&[\AA]     & [\%]
 &  &                 &
{}~[\AA] &            &[eV]   &
                     &\multicolumn{2}{l|}{[\kms]}   \\
\hline
       &     &    &     &    &2717.51  &CrII(007)& 1.48&   M&      &
                                  \\
       &     &    &     &    &2717.525 &MnII(033)& 3.69&   M&      &
                                  \\
2717.70&     &  87&     & 2  &2717.533 &FeII(032)& 0.23&   P&    18&
                                  \\
\cline{6-6}
2719.05&     &  92&     & 3  &2719.027 &Fe I(005)& 0.00&   P&     3& not in
$\alpha$~Lep; IS ?                    \\
2719.90&     &  79&     & 1  &2719.736 &MnII(033)& 3.69&   P&    18&
                                  \\
\cline{6-6}
2721.00&     &  77&     & 3  &2720.902 &Fe I(005)& 0.05&   P&    11&
                                  \\
\hline
\end{tabular}}
\end{small}
\end{table*}

\clearpage
\begin{table*}
\addtocounter{table}{-1}
\caption{continued}
\begin{small}
\centerline{\begin{tabular}{|llllllllllp{24mm}|}
\hline
$\lambda_{\rm obs}$&$W$        &$D$      &$A$
&$Q$                &
$\lambda_{\rm lab}$&ion(mult) &$\chi$ &$C$
&$v$                &remark\\
{}~[\AA]&[\AA]     & [\%]
 &  &                 &
{}~[\AA] &            &[eV]   &
                     &\multicolumn{2}{l|}{[\kms]}   \\
\hline
       &     &    &     &    &2748.98  &CrII(006)& 1.49&   M&      &
                                        \\
       &     &    &     &    &2749.178 &FeII(063)& 1.07&   S&      &
                                        \\
2749.46& 2.08& 100& 1.90& 2  &2749.324 &FeII(062)& 1.04&   P&    15&
                                        \\
       &     &    &     &    &2749.482 &FeII(063)& 1.09&   M&      &
                                        \\
       &     &    &     &    &2750.003 &FeII(199)& 3.19&   M&      &
                                        \\
       &     &    &     &    &2750.140 &Fe I(005)& 0.05&   M&      &
                                        \\
\hline
\end{tabular}}
\end{small}
\end{table*}

\clearpage
\begin{table*}
\addtocounter{table}{-1}
\caption{continued}
\begin{small}
\centerline{\begin{tabular}{|llllllllllp{24mm}|}
\hline
$\lambda_{\rm obs}$&$W$        &$D$      &$A$
&$Q$                &
$\lambda_{\rm lab}$&ion(mult) &$\chi$ &$C$
&$v$                &remark\\
{}~[\AA]&[\AA]     & [\%]
 &  &                 &
{}~[\AA] &            &[eV]   &
                     &\multicolumn{2}{l|}{[\kms]}   \\
\hline
       &     &    &     &    &2779.302 &FeII(234)& 3.25&   S&      &
                                        \\
2779.56& 1.13&  87& 1.17& 3  &2779.832 &Mg I(006)& 2.70&   P&   -28& high
velocity                                      \\
\cline{6-6}
       &     &    &     &    &2780.89  &CrII(058)& 3.74&   M&      &
                                        \\
       &     &    &     &    &2781.418 &Mg I(007)& 2.70&   S&      &
                                        \\
2781.31& 1.12&  69& B B & 1  &2781.418 &Mg I(006)& 2.70&   P&   -11&
                                        \\
\cline{6-6}
2782.71& 0.56&  63& R B & 1  &2782.974 &Mg I(006)& 2.70&   P&   -27&
                                        \\
\hline
\end{tabular}}
\end{small}
\end{table*}

\clearpage
\begin{table*}
\addtocounter{table}{-1}
\caption{continued}
\begin{small}
\centerline{\begin{tabular}{|llllllllllp{24mm}|}
\hline
$\lambda_{\rm obs}$&$W$        &$D$      &$A$
&$Q$                &
$\lambda_{\rm lab}$&ion(mult) &$\chi$ &$C$
&$v$                &remark\\
{}~[\AA]&[\AA]     & [\%]
 &  &                 &
{}~[\AA] &            &[eV]   &
                     &\multicolumn{2}{l|}{[\kms]}   \\
\hline
2834.61& 0.52&  56& 1.20& 2  &         &         &     &   P&      & U; Fe I
                                        \\
\cline{6-6}
2835.91& 1.32&  95& 1.78& 3  &2835.63  &CrII(005)& 1.54&   P&    30& double
peaked                                      \\
       &     &    &     &    &2835.716 &FeII(216)& 3.18&   S&      &
                                        \\
\cline{6-6}
2836.56&     &  66&     & 1  &         &         &     &   P&      & U
                                        \\
\cline{6-6}
2837.61& 0.70&  58& 0.91& 2  &2837.300 &FeII(231)& 3.25&   P&    33&
                                        \\
\cline{6-6}
2838.41&     &  48&     & 1  &2838.120 &Fe I(044)& 0.99&   P&    31&
                                        \\
\hline
\end{tabular}}
\end{small}
\end{table*}

\clearpage
\begin{table*}
\addtocounter{table}{-1}
\caption{continued}
\begin{small}
\centerline{\begin{tabular}{|llllllllllp{24mm}|}
\hline
$\lambda_{\rm obs}$&$W$        &$D$      &$A$
&$Q$                &
$\lambda_{\rm lab}$&ion(mult) &$\chi$ &$C$
&$v$                &remark\\
{}~[\AA]&[\AA]     & [\%]
 &  &                 &
{}~[\AA] &            &[eV]   &
                     &\multicolumn{2}{l|}{[\kms]}   \\
\hline
2872.31& 0.70&  74& 1.14& 3$\ast$&2872.382 &FeII(230)& 3.25&   P&    -7&
                                   \\
\cline{6-6}
2873.92& 1.40&  92& 0.88& 1  &2873.46  &CrII(005)& 1.49&   P&    48&
                                   \\
       &     &    &     &    &2873.81  &CrII(011)& 2.42&   M&      &
                                   \\
\cline{6-6}
       &     &    &     &    &2875.97  &CrII(011)& 2.47&   M&      &
                                   \\
2876.32& 1.90&  97& 1.55& 2  &2876.24  &CrII(005)& 1.50&   P&     8&
                                   \\
       &     &    &     &    &2876.30  &CrII(288)& 4.96&   M&      &
                                   \\
\hline
\end{tabular}}
\end{small}
\end{table*}

\clearpage
\begin{table*}
\addtocounter{table}{-1}
\caption{continued}
\begin{small}
\centerline{\begin{tabular}{|llllllllllp{24mm}|}
\hline
$\lambda_{\rm obs}$&$W$        &$D$      &$A$
&$Q$                &
$\lambda_{\rm lab}$&ion(mult) &$\chi$ &$C$
&$v$                &remark\\
{}~[\AA]&[\AA]     & [\%]
 &  &                 &
{}~[\AA] &            &[eV]   &
                     &\multicolumn{2}{l|}{[\kms]}   \\
\hline
       &     &    &     &    &2917.087 &FeII(336)& 4.48&   M&      &
                                   \\
       &     &    &     &    &2917.365 &V II(011)& 0.33&   M&      &
                                   \\
2917.62& 0.75&  87& 1.07& 3  &2917.465 &FeII(061)& 1.04&   P&    16&
                                   \\
\cline{6-6}
       &     &    &     &    &2919.989 &V II(011)& 0.37&   M&      &
                                   \\
       &     &    &     &    &2920.377 &V II(011)& 0.35&   S&      &
                                   \\
2920.47& 0.74&  84& 1.64& 1  &2920.377 &V II(010)& 0.33&   P&    10&
                                   \\
\hline
\end{tabular}}
\end{small}
\end{table*}

\clearpage
\begin{table*}
\addtocounter{table}{-1}
\caption{continued}
\begin{small}
\centerline{\begin{tabular}{|llllllllllp{24mm}|}
\hline
$\lambda_{\rm obs}$&$W$        &$D$      &$A$
&$Q$                &
$\lambda_{\rm lab}$&ion(mult) &$\chi$ &$C$
&$v$                &remark\\
{}~[\AA]&[\AA]     & [\%]
 &  &                 &
{}~[\AA] &            &[eV]   &
                     &\multicolumn{2}{l|}{[\kms]}   \\
\hline
2965.12&     &  94&     & 1  &2965.036 &FeII(078)& 1.69&   P&     8&
                                   \\
\cline{6-6}
2966.37&     &  68&     & 3  &         &         &     &   P&      & U
                                   \\
\cline{6-6}
2967.02& 0.76&  84& 0.89& 3  &2966.901 &Fe I(001)& 0.00&   P&    12&
                                   \\
\cline{6-6}
2968.57& 0.56&  81& R B & 3  &         &         &     &   P&      & U
                                   \\
\cline{6-6}
2969.12&     &  56&     & 1  &2968.738 &FeII(253)& 3.37&   P&    39&
                                   \\
\cline{6-6}
       &     &    &     &    &2970.106 &Fe I(001)& 0.11&   M&      &
                                   \\
\hline
\end{tabular}}
\end{small}
\end{table*}

\hfill

\end{document}